\documentclass[aps,prb,reprint,amssymb,amsmath,superscriptaddress]{revtex4-1}

\usepackage{mathtext}
\usepackage{amssymb}
\usepackage{graphicx}
\usepackage{amsmath}

\begin{document}

\title{Low magnetic field anomaly of the Hall effect in disordered 2D systems:
Interplay between weak localization and  electron-electron interaction}

\author{G.~M.~Minkov}
\affiliation{Institute of Metal Physics RAS, 620219 Ekaterinburg,
Russia}

\affiliation{Ural State University, 620083 Ekaterinburg, Russia}

\author{A.~V.~Germanenko}
\affiliation{Ural State University, 620083 Ekaterinburg, Russia}

\author{O.~E.~Rut}
\affiliation{Ural State University, 620083 Ekaterinburg, Russia}

\author{A.~A.~Sherstobitov}
\affiliation{Institute of Metal Physics RAS, 620219 Ekaterinburg,
Russia} \affiliation{Ural State University, 620083 Ekaterinburg,
Russia}

\author{B.~N.~Zvonkov}
\affiliation{Physical-Technical Research Institute, University of
Nizhni Novgorod, 603600 Nizhni Novgorod, Russia}

\date{\today}

\begin{abstract}
The nonlinear behavior of the Hall resistivity at low magnetic fields
in single quantum well GaAs/In$_x$Ga$_{1-x}$As/GaAs heterostructures
with degenerated electron gas is studied. It has been found that this
anomaly is accompanied by the weaker temperature dependence of the
conductivity as compared with that predicted by the first-order theory
of the quantum corrections to the conductivity. We show that both
effects  in strongly disordered systems stem from the second order
quantum correction caused by the effect of weak localization on the
interaction correction and vice versa. This correction contributes
mainly to the diagonal component of the conductivity tensor, it depends
on the magnetic field like the weak localization correction and on the
temperature like the interaction contribution.
\end{abstract}

\maketitle

\section{Introduction}
\label{sec:intr}

The quantum corrections to the conductivity, namely the interference or
weak localization (WL) correction and correction due to
electron-electron ({\it e-e}) interaction, wholly determine the
temperature and magnetic field dependences of the conductivity
($\sigma$) at $T\ll E_F, \tau^{-1}$, where $E_F$ and $\tau$ are the
Fermi energy and the transport relaxation time, respectively (hereafter
we set $k_B=\hbar=1$ for brevity).\footnote{We consider the effects
caused by the Landau quantization as being beyond the topic of this
paper.} The modern theory being elaborated since 1980\cite{AA85,
Alei99,Zala01,Gor03,Gor04} allows ones to describe most of experimental
results obtained on the well controlled semiconductor two-dimensional
systems quantitatively. However, one peculiarity, namely the low
magnetic field dependence of the Hall coefficient ($R_H$), referred as
beak in what follows, remains a puzzle. The magnetic field scale of the
beak is close to the transport magnetic field $B_{tr} =\hbar/2el^2$,
where $l$ is the mean free path, i.e., close to the field, in which the
main part of the interference correction is suppressed. As a rule, the
Hall coefficient increases in absolute value with the growing magnetic
field, and the magnitude of the beak is close to that of the negative
magnetoresistivity caused by suppression of the weak localization:
$|\delta R_H/R_H|\sim |\delta \sigma^{WL}|/\sigma$. The existence of
low field anomaly in $R_H$ was pointed out in the pioneering papers on
the quantum corrections.\cite{Pool81,Newson87,Tousson88} In the later
papers the anomaly of $R_H$ behavior is not mentioned, although the
beak is observed practically in all the 2D
structures.\cite{Minkov06,PrvKuntsevich,PrvVitalik}

Theories of the weak localization and interaction correction do not
predict any low magnetic field dependence of the Hall coefficient. The
WL theory asserts that the quantum interference renormalizes the
transport relaxation time and, consequently, does not lead to
correction in the Hall coefficient. The {\it e-e} interaction within
the diffusion regime, $T\tau \ll 1$,  contributes to the longitudinal
conductivity $\sigma_{xx}$ only and this correction does not depend on
the magnetic field while the Zeeman splitting is less than the
temperature, $|\textsl{g}|\mu_B B<T$. So, this correction leads to the
temperature dependence of the Hall coefficient, in the magnetic field
$R_H$ remains constant.\footnote{Generally, the simultaneous existence
of the $B$ dependent WL correction and the interaction correction to
$\sigma_{xx}$, which is independent of the magnetic field, should lead
to the low magnetic field dependence of $R_H$. However, this effect is
the next order of smallness. Moreover, the Hall  coefficient should
decrease in magnitude with the $B$ increase if the sign of the
interaction correction is insulating. The experimental behavior is
opposite.} Thus, the origin of the beak in the $B$ dependence of the
Hall coefficient remains enigmatic.

We have analyzed numerous experimental data regarding the low
field anomaly of the Hall coefficient  for more than thirty
GaAs/In$_x$Ga$_{1-x}$As/GaAs and
Al$_x$Ga$_{1-x}$As/GaAs/Al$_x$Ga$_{1-x}$As structures both with
the electron and hole 2D gas with the carrier density from
$1\times 10^{11}$~cm$^{-2}$ to $2\times 10^{12}$~cm$^{-2}$ and
the mobility from $1\times 10^2$~cm$^2$/Vs to $2\times
10^4$~cm$^2$/Vs. We have not found any correlation between the
beak magnitude and such the structure parameters as the
transport and quantum relaxation time, carries density,
spin-orbit interaction strength and so on. It indicates in our
opinion that there is no universal reason for such a behavior
of the Hall coefficient. However, we believe that in strongly
disordered structures in deep diffusion regime the origin of
the beak in the $R_H$~vs~$B$ dependence is clear. In this paper
we show that it comes from the interplay between the weak
localization and interaction effects. This interplay term
contributes to $\sigma_{xx}$ only like the interaction
correction and depends on the magnetic field like the WL
correction.

\section{Experimental details}
\label{sec:exp} The structures investigated were grown by
metal-organic vapor-phase epitaxy on a semiinsulating GaAs
substrate and consist of 0.5-$\mu$m-thick undoped GaAs
epilayer, a In$_{x}$Ga$_{1-x}$As quantum well with  Sn or Si
$\delta$ layer situated in the well center and a $200$~nm cap
layer of undoped GaAs. The samples were mesa etched into
standard Hall bars and then an Al gate electrode was deposited
by thermal evaporation onto the cap layer through a mask.
Varying the gate voltage ($V_g$) we were able to change the
electron density ($n$) and the conductivity of 2D electron gas
in the quantum well. We studied samples prepared from four
wafers with different well width, doping level and well
composition. All the measurements were carried out in  the
Ohmic regime using DC technique. The results obtained were
mostly analogous, therefore we will discuss the results for the
structure 4261 studied more thoroughly. The quantum well width
in this structure is $8$ nm, indium content in the quantum well
is $0.2$ and tin density in $\delta$ layer is about $2\times
10^{12}$~cm$^{-2}$.

\begin{figure}
\includegraphics[width=\linewidth,clip=true]{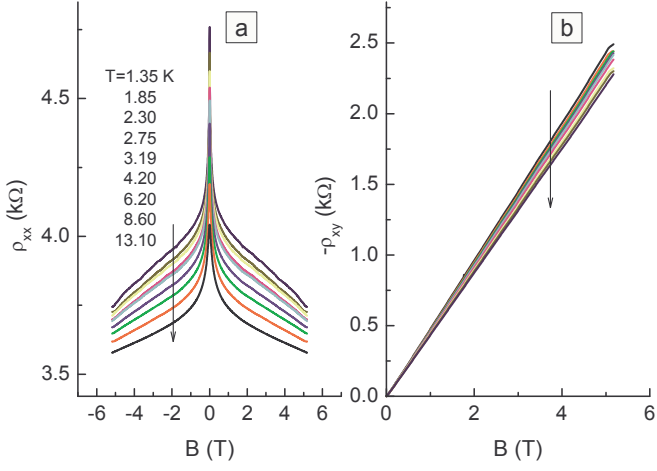}
\caption{(Color online) The magnetic field dependences of $\rho_{xx}$ (a) and $\rho_{xy}$ (b)
for $V_g=-1$~V taken at different temperatures. The arrows indicate the temperature growth.
}\label{f1}
\end{figure}

\section{Results and discussion}
\label{sec:dis}

\begin{figure}
\includegraphics[width=\linewidth,clip=true]{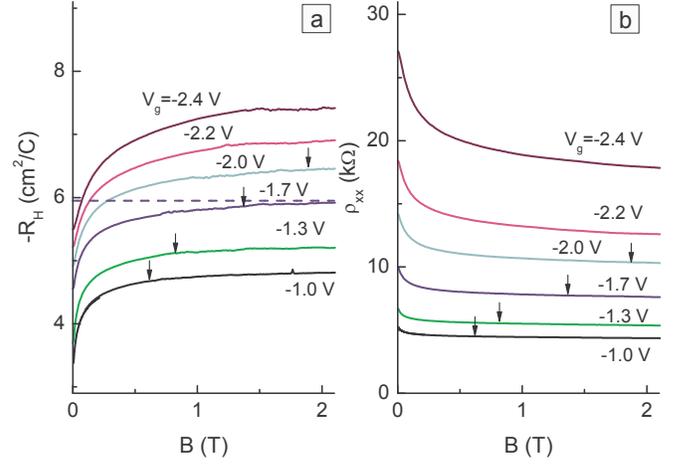}
\caption{(Color online) The magnetic field dependences
of $R_{H}$ (a) and $\rho_{xx}$ (b)
taken  for different gate voltages at $T=1.4$~K.
The dashed line is the Hall constant for $V_g=-1.7$~V obtained from
the linear interpolation of $\rho_{xy}$ within the low magnetic
field domain, $|B|\lesssim 5B_{tr}$.
The arrows indicates the $B_{tr}$ values.}\label{f2}
\end{figure}

Let us consider the magnetic field dependences of $\rho_{xx}$
and $\rho_{xy}$  for  $V_g=-1$~V (Fig.~\ref{f1}) taken at
different temperatures. These dependences are typical for such
a type of systems. The sharp negative magnetoresistance at low
magnetic field [Fig.~\ref{f1}(a)] results from suppression of
the WL contribution. A crossover to the parabolic-like behavior
of $\rho_{xx}$ at $B\gtrsim 2$~T and the decrease of
$\rho_{xy}$ with the temperature increase come from the {\it
e-e} interaction correction. For the first sight, $\rho_{xy}$
linearly depends on the magnetic field [Fig.~\ref{f1}(b)] as
predicted theoretically. Let us, however, inspect the Hall
coefficient, $R_H=\rho_{xy}/B$, which magnetic field
dependences taken for different gate voltages at $T=1.4$~K are
plotted in Fig.~\ref{f2}(a). It is evident that $R_H$ decreases
in magnitude when $B$ goes to zero for all the gate voltages.
Comparing these dependences with that for magnetoresistance
[presented in Fig.~\ref{f2}(b)], one can see that the
characteristic scales in $B$ domain for the $R_H$ beak and for
the interference induced negative magnetoresistance are close;
the main changes happen at $B\lesssim B_{tr}$ in both cases.
Therefore, before to discuss the low field peculiarity of the
Hall coefficient let us analyze the contributions of the
interference and interaction.

First we remind the reader of the basic results of the quantum
correction theory that will be used for analysis. The expression for
the conductivity tensor components taking into account the first order
in $\delta\sigma/\sigma$ corrections are the
following:\footnote{Analyzing the data we will neglect the interaction
corrections in the Cooper channel. Two terms in low magnetic fields
contribute to the low-magnetic  field magnetoconductance. They are
Maki-Thomson correction and correction to the density of states (DoS).
The role  of these terms in the low field magnetoconductivity is
thoroughly considered in Ref.~\onlinecite{Min04-2}. We mention only
that the  DoS term contributes  to $\sigma_{xx}$ and do not to
$\sigma_{xy}$  as well as AA correction, but in contrast to it the  DoS
term yields the magnetoconductivity close in the shape to that  of
interference induced magnetoconductivity. However, our  estimations
show that the DoS correction is three-to-five times smaller in
magnitude than the second-order corrections under our experimental
conditions. They results in about $5-7$\,\% depth of  the beak in the
Hall effect instead of $\sim 30\,$\% observed experimentally.}
\begin{eqnarray}
\sigma_{xx}(B,T)&=&\frac{en\mu(B,T)}{1+\left[\mu(B,T)B\right]^2}+\delta\sigma_{xx}^{ee}(T), \label{eq10} \\
\sigma_{xy}(B,T)&=&\frac{en\mu(B,T)^2B}{1+\left[\mu(B,T)B\right]^2}. \label{eq20}
\end{eqnarray}
In the actual case of $T\tau\ll 1$, the correction
$\delta\sigma_{xx}^{ee}$ is just the Altshuler-Aronov (AA) correction
given
by\cite{AA85,Finkelstein83,Finkelstein84,Cast84-1,Cast84-2,Cast98}
\begin{equation}
 \delta\sigma^{\text{AA}}(T)=K_{ee}^{\text{AA}}G_0\ln(T\tau),
  \label{eq25}
\end{equation}
where
\begin{equation}
 K_{ee}^{\text{AA}}=1+3\left[1-\frac{1+\gamma_2}{\gamma_2}\ln\left(1+\gamma_2\right)\right]
   \label{eq27}
\end{equation}
with $G_0=e^2/\pi h$ and $\gamma_2$ standing for the Landau's
Fermi liquid amplitude. Because the WL correction is reduced to
the renormalization of the transport relaxation
time,\cite{Dmit97} it is incorporated in Eqs.~(\ref{eq10}) and
(\ref{eq20}) into the mobility $\mu$ in such a way that
\begin{equation}
\delta\sigma^{WL}(B,T)=e\,n\,\delta\mu(B,T),
\label{eq30}
\end{equation}
where
\begin{equation}
\delta\sigma^{WL}(0,T)=G_0 \ln{\left[\frac{\tau}{\tau_\phi(T)}\right]},
\label{eq40}
\end{equation}
and
$\Delta\sigma^{WL}(B,T)=\delta\sigma^{WL}(B,T)-\delta\sigma^{WL}(0,T)$
is described by the expression\cite{Hik80,Wit87}
\begin{eqnarray}
\Delta\sigma^{WL}(B,T)&=&\alpha\,G_0{\cal H}\left(\frac{\tau}{\tau_\phi(T)},\frac{B}{B_{tr}}\right), \nonumber \\
{\cal H}(x,y)&=&\psi\left(\frac{1}{2}+\frac{x}{y}\right)-\psi\left(\frac{1}{2}+\frac{1}{y}\right)-\ln{x}.
\label{eq50}
\end{eqnarray}
Here, $\tau_\phi$ is the phase relaxation time,  $\psi(x)$ is a
digamma function, and $\alpha$ is the prefactor, whose value
depends on the conductivity if one takes into account two-loop
localization correction and the interplay of the weak
localization and interaction,\cite{Min04-2}
\begin{equation}
\alpha \simeq 1 - \frac{2\,G_0}{\sigma}, \,\,\,\, \sigma<2\,G_0
\label{eq55}
\end{equation}

\begin{figure}
\includegraphics[width=\linewidth,clip=true]{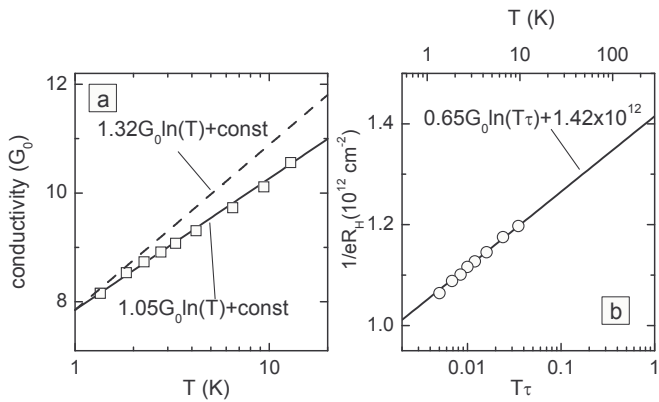}
\caption{(Color online) The temperature dependence of $\sigma$ at $B=0$
(a) and $[eR_H(B=4~\text{T})]^{-1}$ (b) for $V_g=-1.7$~V ($\tau=2.7\times 10^{-14}$~s).
}\label{f3}
\end{figure}

\begin{figure}
\includegraphics[width=\linewidth,clip=true]{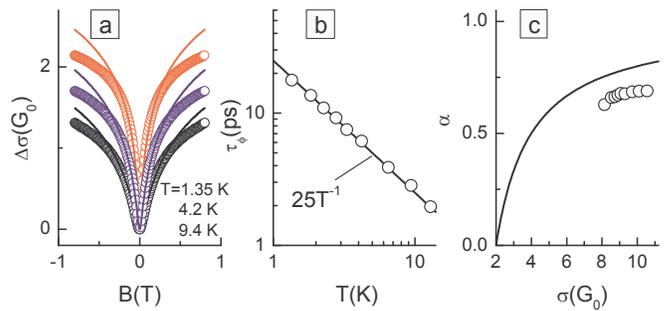}
\caption{(Color online) (a) The magnetic field dependence of $\Delta\sigma$ for
different temperatures. Symbols are the experimental data, lines are the results
of the best fit by Eq.~(\ref{eq50}) within the range of $B$ from $-0.2\,B_{tr}$ to $0.2\,B_{tr}$, where
$B_{tr}=1.26$~T.
(b) The temperature dependences of $\tau_\phi$ found from the fit.
(c) The prefactor $\alpha$ plotted as a function of the conductivity
at $B=0$ driven by the temperature.
The solid line is Eq.~(\ref{eq55}).
}\label{f4}
\end{figure}

We turn now to the analysis of the data. By way of example we consider
the case of $V_g=-1.7$~V. As seen from Fig.~\ref{f3}(a) the temperature
dependence of the conductivity without magnetic field is close to the
logarithmic one, $\sigma(T) = \beta \ln{(T/T_0)}$, with the slope
$\beta$ equal to $1.05\pm 0.05$. To find what portion of the slope
comes from WL let us inspect the low field magnetoconductivity
[Fig.~\ref{f4}(a)]. The electron density $n=(1.42\pm 0.03)\times
10^{12}$~cm$^{-2}$ needed for the analysis we obtain from the
extrapolation of the temperature dependence of the Hall density
$n=1/eR_H$ taken at high magnetic field, $B=4$~T, to $T\tau=1$
[Fig.~\ref{f3}(b)]. Such a dependence of $R_H$ comes from the diffusion
contribution of the interaction, which vanishes at $T\tau=1$. So, the
value of  $1/eR_H$ at $T\tau=1$ actually gives the electron density. An
analysis shows that  Eq.~(\ref{eq50}) the experimental dependences
$\Delta\sigma(B)=1/\rho_{xx}(B)-1/\rho_{xx}(0)$ [see Fig.~\ref{f4}(a)].
The values of $\tau_\phi$ and $\alpha$ found from the fit within
magnetic field range $|B|< 0.2 B_{tr}$ for different temperatures are
plotted in Figs.~\ref{f4}(b) and \ref{f4}(c), respectively. One can see
that $\tau_\phi(T)$ is very close to $1/T$. The prefactor values being
noticeably less than unity, $\alpha=0.6\ldots 0.7$, decreases slightly
with the decreasing conductivity. As Fig.~\ref{f4}(c) shows, such a
behavior agrees well with the theoretical result, Eq.~(\ref{eq55}). So,
the value of $\tau_\phi$ found from the fit of Eq.~(\ref{eq50}) to the
data is the value of the phase breaking time. As mentioned above it is
inversely proportional to the temperature in whole agreement with
theoretical prediction.\cite{AA85} Thus, taking into account
Eq.~(\ref{eq40}) we conclude that the weak localization gives the unit
in the slope of the $\sigma$~vs~$\ln{T}$ dependence at $B=0$.

\begin{figure}
\includegraphics[width=\linewidth,clip=true]{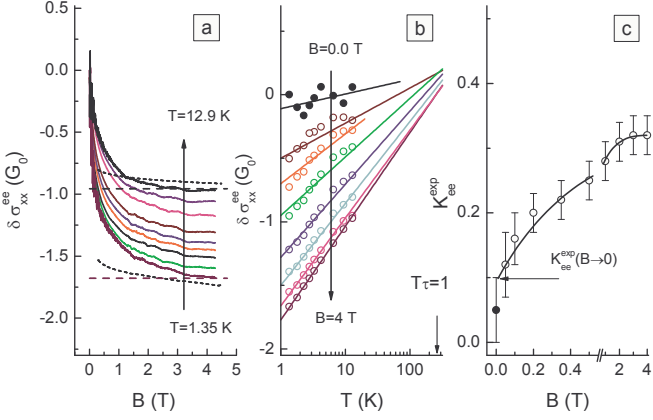}
\caption{(Color online) (a) The magnetic field dependences of the
$\delta\sigma_{xx}^{ee}$ for
different temperatures:  $T=1.35$, $1.8$, $2.27$, $2.77$, $3.3$,
$4.2$, $6.5$, $9.4$, and  $12.9$~K. The dashed lines are
$K_{ee}^\text{AA}\ln{T\tau}$ for $T=1.35$~K and $12.9$~K with $\tau=2.7\times10^{-14}~s$
and $K_{ee}^\text{AA}=0.32$ found
from the $T$ dependence of $R_H$ in high magnetic field [see Fig.~\ref{f3}(b)].
The dotted lines are $\delta\sigma_{xx}^{ee}$ obtained for $T=1.35$~K and $12.9$~K
in the assumption that $R_H$ is independent of $B$.
(b) The temperature dependence of $\delta\sigma_{xx}^{ee}$
for different magnetic field: $B=0$, $0.05$, $0.1$, $0.2$, $0.5$, $1.0$, $2.0$,
and $4.0$~T. Data at $B=0$ are obtained by extrapolation of $\delta\sigma_{xx}^{ee}$~vs~$B$
curves shown in the panel (a) to $B=0$ as described on page \pageref{page10}.
(c) The $B$-dependence of the slope $K_{ee}^{\text{exp}}$ of the
$\delta\sigma_{xx}^{ee}$~vs~$\ln{T}$ dependences shown in the panel (b).
}\label{f5}
\end{figure}

Let us determine now the interaction contribution to the
conductivity. One can find it from the temperature dependence
of the Hall coefficient at high magnetic field [see
Fig.~\ref{f3}(b)] because $\delta R_H/R_H\simeq
-2\delta\sigma_{xx}^{ee}/\sigma_0$ under the condition
$|\delta\sigma_{xx}^{ee}|\ll \sigma_0$. This gives
$\delta\sigma_{xx}^{ee}\simeq 0.32\ln{T\tau}$. However, the
more straightforward way (which does not require the fulfilment
of this condition) is the following.\cite{Min03} Since the
interaction in the diffusion regime contributes to
$\sigma_{xx}$ only, one should find such the contribution to
the conductivity which exists in $\sigma_{xx}$ but is absent in
$\sigma_{xy}$. It can be done by expressing $\mu(B,T)$ from
Eq.~(\ref{eq20}) and substituting it in Eq.(\ref{eq10}). Doing
so we obtain the expression
\begin{equation}
 \delta\sigma_{xx}^{ee}=\frac{1}{\rho_{xx}^{2}+\rho_{xy}^{2}}
 \left[\rho_{xx}-\rho_{xy}\sqrt{\frac{en\left(\rho_{xx}^{2}+
 \rho_{xy}^{2}\right)}{\rho_{xy}B}-1}\right]
 \label{eq57}
\end{equation}
that allows us to find $\delta\sigma_{xx}^{ee}$ using the experimental
quantities $\rho_{xx}$ and $\rho_{xy}$.  The magnetic field dependences
of $\delta\sigma_{xx}^{ee}$ found by this way at different temperatures
are presented in Fig.~\ref{f5}(a). One can see that
$\delta\sigma_{xx}^{ee}$ is practically independent of the magnetic
field while $B\gtrsim (1.5-2)$~T. The temperature dependence of
$\delta\sigma_{xx}^{ee}$ is logarithmic, and the slope
$K_{ee}^{\text{exp}}$ being equal to $0.32\pm 0.05$ remains independent
of the magnetic field at $B\gtrsim 2$~T [see Figs.~\ref{f5}(b) and
\ref{f5}(c)]. Such the behavior agrees well with Eq.~(\ref{eq25}) and,
thus, $K_{ee}^\text{exp}=0.32\pm 0.05$ is just the value of
$K_{ee}^{\text{AA}}$.  Using Eq.~(\ref{eq27}) one obtains
$\gamma_2\simeq0.53$ that is in a good agreement with the results of
Ref.~\onlinecite{Zala01} if one  takes into account the renormalization
effect.\cite{FinRev,Minkov09} Thus, the $T$ dependence of
$\delta\sigma_{xx}^{ee}$ in high magnetic field, $B\gtrsim (1.5-2)$~T,
is determined by the AA quantum correction.

As we have obtained the value of $K_{ee}^{\text{AA}}$ responsible for
the AA contribution, we can compare the value of $\beta$ describing the
experimental $T$ dependence of $\sigma$ at $B=0$ with the value of
$1+K_{ee}^{\text{AA}}$ (recall that $1$ comes from the WL effect) found
from the analysis of the data in the magnetic field. If the model used
and, consequently, Eqs.~(\ref{eq10})--(\ref{eq55}), are correct, the
values of $1+K_{ee}^{\text{AA}}$ and $\beta$ should be equal to each
other. We have that $\beta=1.05\pm 0.05$ [see Fig.~\ref{f3}(a)] is
visibly less than $1+K_{ee}^{\text{AA}}=1.32\pm 0.05$. The reason for
this discrepancy is transparent. The $K_{ee}^{\text{AA}}$ value  has
been above obtained at relatively strong magnetic field, in which the
interaction contribution does not depend on the magnetic field as the
theory predicts. However, inspection of Fig.~\ref{f5} shows that not
only $\delta\sigma_{xx}^{ee}$ diminishes in absolute value at $B\to 0$
but the slope $K_{ee}^{\text{exp}}$ decreases as well. Of course, the
accuracy of $K_{ee}^{\text{exp}}$ determination is not very high in low
magnetic field.  As seen from Fig.~\ref{f5}(a) the experimental
$\delta\sigma_{xx}^{ee}$~vs~$B$ plots are very noisy near $B=0$. The
reason is very clear. The expression, Eq.~(\ref{eq57}), used for the
data treatment contains $B$ in the denominator. Nevertheless,
extrapolating the $K_{ee}^{\text{exp}}$ vs $B$ data to $B=0$ one
obtains $K_{ee}^{\text{exp}}(B\to 0)=0.1\pm 0.05$. Together with $1$
coming from the WL effect we have the value, which practically
coincides with $\beta=1.05\pm 0.05$.

In principle, the $K_{ee}^{\text{exp}}$ change could be induced by the
$K_{ee}^{\text{AA}}$ decrease due to suppression of two of three
triplet channels in Eq.~(\ref{eq25}) due to the Zeeman
effect.\cite{Cast84-1,Cast98,Fin84,Raim90,Minkov07} However, this
effect is negligible in our case due to the low value of the effective
$\textsl{g}$-factor, $\textsl{g}\sim 0.5$. Moreover, if the Zeeman
splitting would be important, the $T$ dependence of
$\delta\sigma_{xx}^{ee}$  in high magnetic field should be strongly
nonlogarithmic as it takes place in 2D hole gas (see Fig.~2 in
Ref.~\onlinecite{Minkov07}).

It is essential to note that the strong decrease of
$\delta\sigma_{xx}^{ee}$ in absolute value with lowering magnetic filed
results from the beak in $R_H$~vs~$B$ dependence. Really, if one uses
the linear interpolation of $\rho_{xy}$ within the range $\pm
(4-5)B_{tr}$ in the above procedure, i.e., one supposes that $R_H$ is
constant as shown in Fig.~\ref{f2}(b) for $V_g=-1.7$~V by the dashed
line, we obtain $\delta\sigma_{xx}^{ee}$, which is practically
independent of the magnetic field [dotted lines in Fig.~\ref{f5}(a)].

\begin{figure}
\includegraphics[width=\linewidth,clip=true]{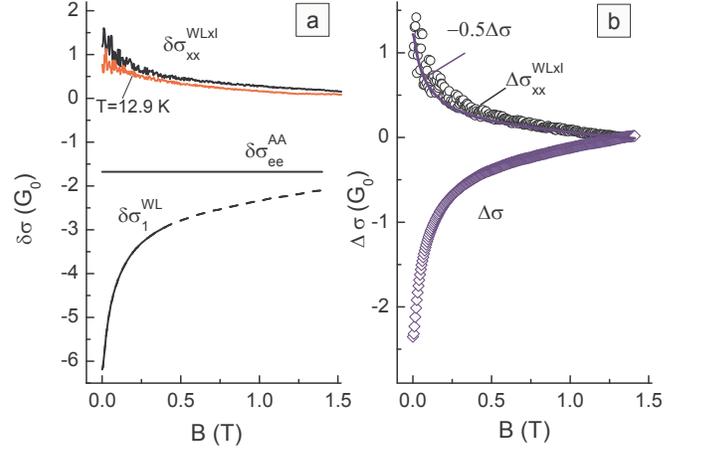}
\caption{(Color online) (a) The magnetic field dependences of
$\delta\sigma^\text{WL}_1$, $\delta\sigma_{ee}^\text{AA}$ and
$\delta\sigma_{xx}^{\text{WL}\times\text{I}}$ for $T=1.35$~K. The dashed line indicates that
the estimate for $\delta\sigma^\text{WL}_1$ is rough
in this range, because Eq.~(\ref{eq60}) is valid when $B\ll B_{tr}$ (b) The
experimental dependences
$\Delta\sigma=\rho_{xx}^{-1}(B)-\rho_{xx}^{-1}(B_{tr})$ and
$\Delta\sigma^{\text{WL}\times\text{I}}_{xx}=\delta\sigma^{\text{WL}\times\text{I}}_{xx}(B)-
\delta\sigma^{\text{WL}\times\text{I}}_{xx}(B_{tr})$.
}\label{f6}
\end{figure}

Thus, there is common reason behind the beak in the $R_H$~vs~$B$
dependence and the existence of magnetic field dependence of
$\delta\sigma_{xx}^{ee}$. Because the $T$ and $B$ changes of the
conductivity are rather large ($\delta\sigma$ is about $(20-30)$\% of
$\sigma$), it is natural to assume that the second order corrections
play an important role under our conditions.

The role of the second order effects is studied in the number of
paper.\cite{Hik81,Wegner79,Wegner80,Wegner89,Alei99,Baranov02,Min04-2,Punnoose05}
The second-order interaction correction (not involving Cooperons),
$\delta\sigma_2^{\text{I}}$, logarithmically depends on the
temperature, but does not depend on the magnetic field analogously to
the AA correction.\cite{Baranov02,Punnoose05} That is why it barely
gives the correction to $K_{ee}^{\text{AA}}$, Eq.~(\ref{eq27}), and
does not affect the low magnetic field magnetoresistance
$\Delta\sigma(B)$. The other two second order terms have an impact on
$\Delta\sigma(B)$. They are $\delta\sigma^{\text{WL}\times\text{I}}$
coming from the interplay between the weak localization and the
interaction effects,\cite{Alei99} and $\delta\sigma_2^{\text{WL}}$,
which is the second order interference correction.\cite{Min04-2} Except
for opposite sign the magnetic field dependences of both terms are
close to that for the first order interference correction. Namely this
fact results in the appearance of $\alpha<1$ in Eq.~(\ref{eq50}).
Science the interference correction stems from the ($B$-dependent)
correction to the impurity scattering cross section and hence
renormalizes the value of the elastic transport scattering rate
$1/\tau$, the higher order interference corrections  do not contribute
to the Hall effect\cite{Shapiro81,PrvIgor} analogously to the first
order one.\cite{AA85} Moreover, $\delta\sigma_2^{\text{WL}}$ does not
contribute to the $T$ dependence of $\sigma$ at zero magnetic field,
since the terms of the second and third orders cancel out in the
interference correction.\cite{Hik81,Wegner79} Thus, it is reasonable to
assume that the main effect comes from the interplay term
$\delta\sigma^{\text{WL}\times\text{I}}$.

Generally, the interplay effect may give corrections to both components
of the conductivity tensor.  We designate them as
$\delta\sigma_{xx}^{\text{WL}\times\text{I}}$ and
$\delta\sigma_{xy}^{\text{WL}\times\text{I}}$. Because
$\sigma_{xy}^{\text{WL}\times\text{I}}=0$ at $B=0$, the difference
between $\beta$ and $1+K_{ee}^\text{AA}$ results from
$\delta\sigma_{xx}^{\text{WL}\times\text{I}}$. Suppose that
$\delta\sigma_{xy}^{\text{WL}\times\text{I}}$ is small in the presence
of magnetic field as well:
$\delta\sigma_{xy}^{\text{WL}\times\text{I}}\ll\mu B\,
\delta\sigma_{xx}^{\text{WL}\times\text{I}}$. In this case the quantity
$\delta\sigma_{xx}^{ee}$ found above is just the sum of the AA
correction, which is independent of the magnetic field and
logarithmically dependent on the temperature, and the second order
correction $\delta\sigma_{xx}^{\text{WL}\times\text{I}}$, which depends
both on $T$ and $B$:
$\delta\sigma_{xx}^{ee}=\delta\sigma_{ee}^\text{AA}(T)+\delta\sigma_{xx}^
{\text{WL}\times\text{I}}(B,T)$.
In Fig.~\ref{f6}(a), the value of
$\delta\sigma_{xx}^{\text{WL}\times\text{I}}$ found as
$\delta\sigma_{xx}^{ee}-K_{ee}^\text{AA}\ln{T\tau}$ with
$\tau=2.7\times 10^{-14}$~s and  $K_{ee}^\text{AA}=0.32$ are plotted
against the magnetic field. For comparison, the one-loop WL corrections
$\delta\sigma_{ee}^\text{AA}(T)$ and $\delta\sigma^{\text{WL}}_1$ are
depicted in the same figure. The $\delta\sigma^{\text{WL}}_1$
correction is obtained from the experimental data in accordance with
Eqs.~(\ref{eq40}) and (\ref{eq50}) as follows
\begin{equation}
\delta\sigma^{\text{WL}}_1\simeq\frac{\Delta{\sigma(B)}}
{\alpha}-\ln{\left(\frac{\tau_\phi}{\tau}\right)}. \label{eq60}
\end{equation}
Two important properties of
$\delta\sigma_{xx}^{\text{WL}\times\text{I}}$ are evident.  First, the
interplay correction is metallic-like in contrast to the WL and AA
corrections, i.e., it increases with temperature decrease [see
Fig.~\ref{f6}(a)]. Qualitatively, this explains the difference between
$1+K_{ee}^\text{AA}$ and $\beta$. Second,  the $B$ range, where the
main changes in $\delta\sigma_{xx}^{\text{WL}\times\text{I}}$ occur, is
the same as for $\delta\sigma^{\text{WL}}_1$: $B< B_{tr}\simeq 1.36$~T.
The fact that $\delta\sigma_{xx}^{\text{WL}\times\text{I}}$~vs~$B$
curve is close in the shape to the low magnetic field
magnetoconductance is illustrated in Fig.~\ref{f6}(b), where the
dependences $\Delta\sigma(B)=\rho_{xx}^{-1}(B)-\rho_{xx}^{-1}(B_{tr})$
and
$\Delta\sigma^{\text{WL}\times\text{I}}_{xx}(B)=\delta\sigma^
{\text{WL}\times\text{I}}_{xx}(B)-\delta\sigma^{\text{WL}\times\text{I}}_{xx}(B_{tr})$
are shown. As seen $\Delta\sigma(B)$  multiplied by the factor $\gamma$
of $-0.5$ fits the $\Delta\sigma^{\text{WL}\times\text{I}}_{xx}$ dots
rather well.

The second property has been used to obtain the value of
$\delta\sigma_{xx}^{\text{WL}\times\text{I}}(B=0)$ [and
$\delta\sigma_{xx}^{ee}(B=0)$ shown in Fig.~\ref{f5}(b) by solid
circles\label{page10}]. We have interpolated the experimental $B$
dependences of $\delta\sigma_{xx}^{\text{WL}\times\text{I}}$ (and
$\delta\sigma_{xx}^{ee}$) at $B<B_{tr}$ by the experimental curve
$\Delta\sigma(B)$ excluding the noisy vicinity of $B=0$ and
interpolated it to $B=0$. The $T$ dependence of
$\delta\sigma_{xx}^{\text{WL}\times\text{I}}(0)$ found in such a way is
close to the logarithmic one with the slope
$K^{\text{WL}\times\text{I}}= -0.3\pm 0.1$.

\begin{figure}
\includegraphics[width=\linewidth,clip=true]{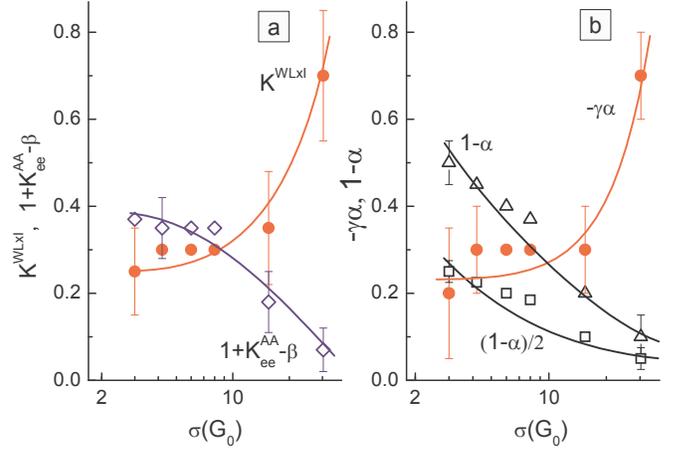}
\caption{(Color online) (a) The values of $K^{\text{WL}\times\text{I}}$ (circles)
and the difference between $1+K_{ee}^\text{AA}$ and $\beta$ (diamonds) as a function of
the conductivity at $T=1.35$~K.
(b) The conductivity dependence of $-\gamma\, \alpha$ (circles), $1-\alpha$ (triangles), and
$(1-\alpha)/2$ (squares).
Lines are provided as a guide to the eye. The conductivity
in both panels is driven by the gate voltage.} \label{f7}
\end{figure}

Thus, the temperature dependence of the conductivity at $B=0$ caused by
all three contributions $\delta\sigma^\text{WL}_1$,
$\delta\sigma_{ee}^\text{AA}$, and
$\delta\sigma_{xx}^{\text{WL}\times\text{I}}$  is very close to that
observed experimentally. The total slope equal to
$1+K_{ee}^\text{AA}+K^{\text{WL}\times\text{I}}= 1.02\pm 0.1$ is in a
good agreement with the experimental value $\beta=1.05\pm 0.05$ [see
Fig.~\ref{f3}(a)].

Analysis described has been performed within wide range of the
conductivity driven by the gate voltage. The values of
$K^{\text{WL}\times\text{I}}$ plotted against $\sigma$ at $T=1.35$~K
are shown in Fig.~\ref{f7}(a). In the same figure the difference
between $1+K_{ee}^\text{AA}$ and  $\beta$ is depicted. It is seen that
both data are close to each other at low conductivity,
$\sigma<20\,G_0$. At higher conductivity, they diverge drastically.

The contribution of $\delta\sigma_{xx}^{\text{WL}\times\text{I}}$ to
the magnetoconductivity is illustrated by Fig.~\ref{f7}(b). We
characterize it by the product $\gamma\,\alpha$ which is
$\Delta\sigma_{xx}^{\text{WL}\times\text{I}}(B)$ to
$\Delta\sigma^\text{WL}_1(B)$ ratio. If one supposes that
$\delta\sigma_{xx}^{\text{WL}\times\text{I}}$ is alone and there is no
$\Delta\sigma^\text{WL}_2$ contribution to the magnetoconductivity,
this value should be equal to $1-\alpha$. If these correction are the
same in magnitude (as it turns out theoretically for the short-range
interaction\cite{Min04-2}), the $\gamma\,\alpha$ value has to be equal
to half of this value. As seen from Fig.~\ref{f7}(b) the agreement is
satisfactory with both the cases at $\sigma < (10-15)\,G_0$ if one
takes into account the experimental error. At $\sigma \simeq 30\,G_0$
the difference becomes crucial.

The discrepancy between the data obtained in different manner evident
in Figs.~\ref{f7}(a) and \ref{f7}(b) at high conductivity probably
means that our assumption about smallness of the correction to the Hall
conductivity $\sigma_{xy}$ is not valid in this case suggesting further
investigations are needed to understand the origin of the low field
anomaly in the Hall effect in the relatively clean systems.

Thus, the second order correction
$\delta\sigma_{xx}^{\text{WL}\times\text{I}}$ caused by interplay
between the WL and interaction corrections is of importance in our
case. At low conductivity, $\sigma< (10-15)\,G_0$, this correction
contributes to the diagonal component of the conductivity tensor
$\sigma_{xx}$ only. Its temperature dependence is metallic like in
contrast to the WL and AA corrections which are insulating. Its
magnetic field dependence is close in the shape to that of the WL
correction, although the magnetoconductivity itself is negative in
contrast to that induced by suppression of the weak localization.
Existence of this correction results in: (i) the depressing of the
interference induced low magnetic field magnetoresistance; (ii) the
difference between the slope of the $\sigma$~vs~$\ln{T}$ dependence and
the value of $1+K_{ee}^\text{AA}$; (iii) the occurrence of the beak in
the $R_H$~vs~$B$ dependence in low magnetic field.

\section{Conclusion}
We have studied the nonlinear behavior of the Hall resistivity in the
vicinity of zero magnetic field. Investigating the two-dimensional
electron gas in strongly disordered GaAs/In$_x$Ga$_{1-x}$As/GaAs
quantum well  we have shown that the anomaly of the Hall resistance and
impossibility of description of the temperature dependences of
zero-field conductivity by taking into account only two first order WL
and AA quantum corrections are explained by significant contribution of
the second order correction resulting from the effect of weak
localization on the interaction correction and vice versa in disordered
systems with $\sigma<(10-15)\,G_0$. The experimental results are
satisfactorily interpreted under assumption that this correction
contributes to the diagonal component of the conductivity tensor
$\sigma_{xx}$ only, its magnetic field dependence is close to that of
the weak localization correction, although the temperature dependence
is metallic-like.

\subsection*{Acknowledgments}
We would like to thank I.~S.~Burmistrov and  I.~V.~Gornyi for
illuminating discussions. This work has been supported in part by the
RFBR (Grant Nos 08-02-00662, 09-02-00789, and 10-02-00481).

%\bibliography{QuantumCorrections}

\begin{thebibliography}{37}%
\makeatletter
\providecommand \@ifxundefined [1]{%
 \@ifx{#1\undefined}
}%
\providecommand \@ifnum [1]{%
 \ifnum #1\expandafter \@firstoftwo
 \else \expandafter \@secondoftwo
 \fi
}%
\providecommand \@ifx [1]{%
 \ifx #1\expandafter \@firstoftwo
 \else \expandafter \@secondoftwo
 \fi
}%
\providecommand \natexlab [1]{#1}%
\providecommand \enquote  [1]{``#1''}%
\providecommand \bibnamefont  [1]{#1}%
\providecommand \bibfnamefont [1]{#1}%
\providecommand \citenamefont [1]{#1}%
\providecommand \href@noop [0]{\@secondoftwo}%
\providecommand \href [0]{\begingroup \@sanitize@url \@href}%
\providecommand \@href[1]{\@@startlink{#1}\@@href}%
\providecommand \@@href[1]{\endgroup#1\@@endlink}%
\providecommand \@sanitize@url [0]{\catcode `\\12\catcode `\$12\catcode
  `\&12\catcode `\#12\catcode `\^12\catcode `\_12\catcode `\%12\relax}%
\providecommand \@@startlink[1]{}%
\providecommand \@@endlink[0]{}%
\providecommand \url  [0]{\begingroup\@sanitize@url \@url }%
\providecommand \@url [1]{\endgroup\@href {#1}{\urlprefix }}%
\providecommand \urlprefix  [0]{URL }%
\providecommand \Eprint [0]{\href }%
\@ifxundefined \urlstyle {%
  \providecommand \doi  [0]{\begingroup \@sanitize@url \@doi}%
  \providecommand \@doi [1]{\endgroup \@@startlink {\doibase
  #1}doi:\discretionary {}{}{}#1\@@endlink }%
}{%
  \providecommand \doi  [0]{doi:\discretionary{}{}{}\begingroup
  \urlstyle{rm}\Url }%
}%
\providecommand \doibase [0]{http://dx.doi.org/}%
\providecommand \Doi [0]{\begingroup \@sanitize@url \@Doi }%
\providecommand \@Doi  [1]{\endgroup\@@startlink{\doibase#1}\@@Doi}%
\providecommand \@@Doi [1]{#1\@@endlink}%
\providecommand \selectlanguage [0]{\@gobble}%
\providecommand \bibinfo  [0]{\@secondoftwo}%
\providecommand \bibfield  [0]{\@secondoftwo}%
\providecommand \translation [1]{[#1]}%
\providecommand \BibitemOpen [0]{}%
\providecommand \bibitemStop [0]{}%
\providecommand \bibitemNoStop [0]{.\EOS\space}%
\providecommand \EOS [0]{\spacefactor3000\relax}%
\providecommand \BibitemShut  [1]{\csname bibitem#1\endcsname}%
%</preamble>
\bibitem [{Note1()}]{Note1}%
  \BibitemOpen
  \bibinfo {note} {We consider the effects caused by the Landau quantization as
  being beyond the topic of this paper.}\BibitemShut {Stop}%
\bibitem [{\citenamefont {Altshuler}\ and\ \citenamefont
  {Aronov}(1985)}]{AA85}%
  \BibitemOpen
  \bibfield  {author} {\bibinfo {author} {\bibfnamefont {B.~L.}\ \bibnamefont
  {Altshuler}}\ and\ \bibinfo {author} {\bibfnamefont {A.~G.}\ \bibnamefont
  {Aronov}},\ }in\ \href@noop {} {\emph {\bibinfo {booktitle}
  {Electron-Electron Interaction in Disordered Systems}}},\ \bibinfo {editor}
  {edited by\ \bibinfo {editor} {\bibfnamefont {A.~L.}\ \bibnamefont {Efros}}\
  and\ \bibinfo {editor} {\bibfnamefont {M.}~\bibnamefont {Pollak}}}\ (\bibinfo
   {publisher} {North Holland},\ \bibinfo {address} {Amsterdam},\ \bibinfo
  {year} {1985})\ p.~\bibinfo {pages} {1}\BibitemShut {NoStop}%
\bibitem [{\citenamefont {Aleiner}\ \emph {et~al.}(1999)\citenamefont
  {Aleiner}, \citenamefont {Altshuler},\ and\ \citenamefont
  {E.Gershenzon}}]{Alei99}%
  \BibitemOpen
  \bibfield  {author} {\bibinfo {author} {\bibfnamefont {I.~L.}\ \bibnamefont
  {Aleiner}}, \bibinfo {author} {\bibfnamefont {B.~L.}\ \bibnamefont
  {Altshuler}}, \ and\ \bibinfo {author} {\bibfnamefont {M.}~\bibnamefont
  {E.Gershenzon}},\ }\href@noop {} {\bibfield  {journal} {\bibinfo  {journal}
  {Waves Random Media},\ }\textbf {\bibinfo {volume} {9}},\ \bibinfo {pages}
  {201 } (\bibinfo {year} {1999})}\BibitemShut {NoStop}%
\bibitem [{\citenamefont {Zala}\ \emph {et~al.}(2001)\citenamefont
    {Zala},
  \citenamefont {Narozhny},\ and\ \citenamefont {Aleiner}}]{Zala01}%
  \BibitemOpen
  \bibfield  {author} {\bibinfo {author} {\bibfnamefont {G.}~\bibnamefont
  {Zala}}, \bibinfo {author} {\bibfnamefont {B.~N.}\ \bibnamefont {Narozhny}},
  \ and\ \bibinfo {author} {\bibfnamefont {I.~L.}\ \bibnamefont {Aleiner}},\
  }\Doi {10.1103/PhysRevB.64.214204} {\bibfield  {journal} {\bibinfo  {journal}
  {Phys. Rev. B},\ }\textbf {\bibinfo {volume} {64}},\ \bibinfo {pages}
  {214204} (\bibinfo {year} {2001})}\BibitemShut {NoStop}%
\bibitem [{\citenamefont {Gornyi}\ and\ \citenamefont {Mirlin}(2003)}]{Gor03}%
  \BibitemOpen
  \bibfield  {author} {\bibinfo {author} {\bibfnamefont {I.~V.}\ \bibnamefont
  {Gornyi}}\ and\ \bibinfo {author} {\bibfnamefont {A.~D.}\ \bibnamefont
  {Mirlin}},\ }\Doi {10.1103/PhysRevLett.90.076801} {\bibfield  {journal}
  {\bibinfo  {journal} {Phys. Rev. Lett.},\ }\textbf {\bibinfo {volume} {90}},\
  \bibinfo {pages} {076801} (\bibinfo {year} {2003})}\BibitemShut {NoStop}%
\bibitem [{\citenamefont {Gornyi}\ and\ \citenamefont {Mirlin}(2004)}]{Gor04}%
  \BibitemOpen
  \bibfield  {author} {\bibinfo {author} {\bibfnamefont {I.~V.}\ \bibnamefont
  {Gornyi}}\ and\ \bibinfo {author} {\bibfnamefont {A.~D.}\ \bibnamefont
  {Mirlin}},\ }\Doi {10.1103/PhysRevB.69.045313} {\bibfield  {journal}
  {\bibinfo  {journal} {Phys. Rev. B},\ }\textbf {\bibinfo {volume} {69}},\
  \bibinfo {pages} {045313} (\bibinfo {year} {2004})}\BibitemShut {NoStop}%
\bibitem [{\citenamefont {Poole}\ \emph {et~al.}(1981)\citenamefont
    {Poole},
  \citenamefont {Pepper},\ and\ \citenamefont {Glew}}]{Pool81}%
  \BibitemOpen
  \bibfield  {author} {\bibinfo {author} {\bibfnamefont {D.~A.}\ \bibnamefont
  {Poole}}, \bibinfo {author} {\bibfnamefont {M.}~\bibnamefont {Pepper}}, \
  and\ \bibinfo {author} {\bibfnamefont {R.~W.}\ \bibnamefont {Glew}},\
  }\href@noop {} {\bibfield  {journal} {\bibinfo  {journal} {J. Phys C: Solid
  State Phys},\ }\textbf {\bibinfo {volume} {14}},\ \bibinfo {pages} {L995 }
  (\bibinfo {year} {1981})}\BibitemShut {NoStop}%
\bibitem [{\citenamefont {Newson}\ \emph {et~al.}(1987)\citenamefont
    {Newson},
  \citenamefont {Pepper}, \citenamefont {Hall},\ and\ \citenamefont
  {Hill}}]{Newson87}%
  \BibitemOpen
  \bibfield  {author} {\bibinfo {author} {\bibfnamefont {D.~J.}\ \bibnamefont
  {Newson}}, \bibinfo {author} {\bibfnamefont {M.}~\bibnamefont {Pepper}},
  \bibinfo {author} {\bibfnamefont {E.~Y.}\ \bibnamefont {Hall}}, \ and\
  \bibinfo {author} {\bibfnamefont {G.}~\bibnamefont {Hill}},\ }\href@noop {}
  {\bibfield  {journal} {\bibinfo  {journal} {J. Phys. C: Solid State Phys},\
  }\textbf {\bibinfo {volume} {20}},\ \bibinfo {pages} {4369 } (\bibinfo {year}
  {1987})}\BibitemShut {NoStop}%
\bibitem [{\citenamefont {Tousson}\ and\ \citenamefont
  {Ovadyahu}(1988)}]{Tousson88}%
  \BibitemOpen
  \bibfield  {author} {\bibinfo {author} {\bibfnamefont {E.}~\bibnamefont
  {Tousson}}\ and\ \bibinfo {author} {\bibfnamefont {Z.}~\bibnamefont
  {Ovadyahu}},\ }\Doi {10.1103/PhysRevB.38.12290} {\bibfield  {journal}
  {\bibinfo  {journal} {Phys. Rev. B},\ }\textbf {\bibinfo {volume} {38}},\
  \bibinfo {pages} {12290} (\bibinfo {year} {1988})}\BibitemShut {NoStop}%
\bibitem [{\citenamefont {Minkov}\ \emph {et~al.}(2006)\citenamefont
    {Minkov},
  \citenamefont {Germanenko}, \citenamefont {Rut}, \citenamefont
  {Sherstobitov}, \citenamefont {Larionova}, \citenamefont {Bakarov},\ and\
  \citenamefont {Zvonkov}}]{Minkov06}%
  \BibitemOpen
  \bibfield  {author} {\bibinfo {author} {\bibfnamefont {G.~M.}\ \bibnamefont
  {Minkov}}, \bibinfo {author} {\bibfnamefont {A.~V.}\ \bibnamefont
  {Germanenko}}, \bibinfo {author} {\bibfnamefont {O.~E.}\ \bibnamefont {Rut}},
  \bibinfo {author} {\bibfnamefont {A.~A.}\ \bibnamefont {Sherstobitov}},
  \bibinfo {author} {\bibfnamefont {V.~A.}\ \bibnamefont {Larionova}}, \bibinfo
  {author} {\bibfnamefont {A.~K.}\ \bibnamefont {Bakarov}}, \ and\ \bibinfo
  {author} {\bibfnamefont {B.~N.}\ \bibnamefont {Zvonkov}},\ }\Doi
  {10.1103/PhysRevB.74.045314} {\bibfield  {journal} {\bibinfo  {journal}
  {Phys. Rev. B},\ }\textbf {\bibinfo {volume} {74}},\ \bibinfo {pages}
  {045314} (\bibinfo {year} {2006})}\BibitemShut {NoStop}%
\bibitem [{\citenamefont {Kuntsevich}\ and\ \citenamefont
  {Pudalov}()}]{PrvKuntsevich}%
  \BibitemOpen
  \bibfield  {author} {\bibinfo {author} {\bibfnamefont {A.~Y.}\ \bibnamefont
  {Kuntsevich}}\ and\ \bibinfo {author} {\bibfnamefont {V.~M.}\ \bibnamefont
  {Pudalov}},\ }\href@noop {} {\bibinfo  {journal} {private
  communication}}\BibitemShut {NoStop}%
\bibitem [{\citenamefont {Tkachenko}\ and\ \citenamefont
  {Kvon}()}]{PrvVitalik}%
  \BibitemOpen
\bibfield  {journal} {  }\bibfield  {author} {\bibinfo {author}
{\bibfnamefont
  {V.~A.}\ \bibnamefont {Tkachenko}}\ and\ \bibinfo {author} {\bibfnamefont
  {Z.~D.}\ \bibnamefont {Kvon}},\ }\href@noop {} {\bibinfo  {journal} {private
  communication}}\BibitemShut {NoStop}%
\bibitem [{Note2()}]{Note2}%
  \BibitemOpen
\bibfield  {journal} {  }\bibinfo {note} {Generally, the simultaneous
existence
  of the $B$ dependent WL correction and the interaction correction to $\sigma
  _{xx}$, which is independent of the magnetic field, should lead to the low
  magnetic field dependence of $R_H$. However, this effect is the next order of
  smallness. Moreover, the Hall coefficient should decrease in magnitude with
  the $B$ increase if the sign of the interaction correction is insulating. The
  experimental behavior is opposite.}\BibitemShut {Stop}%
\bibitem [{Note3()}]{Note3}%
  \BibitemOpen
  \bibinfo {note} {Analyzing the data we will neglect the interaction
  corrections in the Cooper channel. Two terms in low magnetic fields
  contribute to the low-magnetic field magnetoconductance. They are
  Maki-Thomson correction and correction to the density of states (DoS). The
  role of these terms in the low field magnetoconductivity is thoroughly
  considered in Ref.~\protect \rev@citealpnum {Min04-2}. We mention only that
  the DoS term contributes to $\sigma _{xx}$ and do not to $\sigma _{xy}$ as
  well as AA correction, but in contrast to it the DoS term yields the
  magnetoconductivity close in the shape to that of interference induced
  magnetoconductivity. However, our estimations show that the DoS correction is
  three-to-five times smaller in magnitude than the second-order corrections
  under our experimental conditions. They results in about $5-7$\protect
  \tmspace +\thinmuskip {.1667em}\% depth of the beak in the Hall effect
  instead of $\sim 30\protect \tmspace +\thinmuskip {.1667em}$\% observed
  experimentally.}\BibitemShut {Stop}%
\bibitem [{\citenamefont {Finkel'stein}(1983)}]{Finkelstein83}%
  \BibitemOpen
  \bibfield  {author} {\bibinfo {author} {\bibfnamefont {A.~M.}\ \bibnamefont
  {Finkel'stein}},\ }\href@noop {} {\bibfield  {journal} {\bibinfo  {journal}
  {Zh. Eksp. Teor. Fiz.},\ }\textbf {\bibinfo {volume} {84}},\ \bibinfo {pages}
  {168} (\bibinfo {year} {1983})},\ \translation{Sov. Phys. JETP \textbf{57},
  97 (1983)}\BibitemShut {NoStop}%
\bibitem [{\citenamefont {Finkel'stein}(1984){\natexlab{a}}}]{Finkelstein84}%
  \BibitemOpen
  \bibfield  {author} {\bibinfo {author} {\bibfnamefont {A.~M.}\ \bibnamefont
  {Finkel'stein}},\ }\href@noop {} {\bibfield  {journal} {\bibinfo  {journal}
  {Z. Phys. B: Condens. Matter},\ }\textbf {\bibinfo {volume} {56}},\ \bibinfo
  {pages} {189} (\bibinfo {year} {1984}{\natexlab{a}})}\BibitemShut {NoStop}%
\bibitem [{\citenamefont {Castellani}\ \emph
  {et~al.}(1984){\natexlab{a}}\citenamefont {Castellani}, \citenamefont
  {Di~Castro}, \citenamefont {Lee},\ and\ \citenamefont {Ma}}]{Cast84-1}%
  \BibitemOpen
  \bibfield  {author} {\bibinfo {author} {\bibfnamefont {C.}~\bibnamefont
  {Castellani}}, \bibinfo {author} {\bibfnamefont {C.}~\bibnamefont
  {Di~Castro}}, \bibinfo {author} {\bibfnamefont {P.~A.}\ \bibnamefont {Lee}},
  \ and\ \bibinfo {author} {\bibfnamefont {M.}~\bibnamefont {Ma}},\ }\Doi
  {10.1103/PhysRevB.30.527} {\bibfield  {journal} {\bibinfo  {journal} {Phys.
  Rev. B},\ }\textbf {\bibinfo {volume} {30}},\ \bibinfo {pages} {527}
  (\bibinfo {year} {1984}{\natexlab{a}})}\BibitemShut {NoStop}%
\bibitem [{\citenamefont {Castellani}\ \emph
  {et~al.}(1984){\natexlab{b}}\citenamefont {Castellani}, \citenamefont
  {Di~Castro}, \citenamefont {Lee}, \citenamefont {Ma}, \citenamefont
  {Sorella},\ and\ \citenamefont {Tabet}}]{Cast84-2}%
  \BibitemOpen
  \bibfield  {author} {\bibinfo {author} {\bibfnamefont {C.}~\bibnamefont
  {Castellani}}, \bibinfo {author} {\bibfnamefont {C.}~\bibnamefont
  {Di~Castro}}, \bibinfo {author} {\bibfnamefont {P.~A.}\ \bibnamefont {Lee}},
  \bibinfo {author} {\bibfnamefont {M.}~\bibnamefont {Ma}}, \bibinfo {author}
  {\bibfnamefont {S.}~\bibnamefont {Sorella}}, \ and\ \bibinfo {author}
  {\bibfnamefont {E.}~\bibnamefont {Tabet}},\ }\Doi {10.1103/PhysRevB.30.1596}
  {\bibfield  {journal} {\bibinfo  {journal} {Phys. Rev. B},\ }\textbf
  {\bibinfo {volume} {30}},\ \bibinfo {pages} {1596} (\bibinfo {year}
  {1984}{\natexlab{b}})}\BibitemShut {NoStop}%
\bibitem [{\citenamefont {Castellani}\ \emph
    {et~al.}(1998)\citenamefont
  {Castellani}, \citenamefont {Di~Castro},\ and\ \citenamefont {Lee}}]{Cast98}%
  \BibitemOpen
  \bibfield  {author} {\bibinfo {author} {\bibfnamefont {C.}~\bibnamefont
  {Castellani}}, \bibinfo {author} {\bibfnamefont {C.}~\bibnamefont
  {Di~Castro}}, \ and\ \bibinfo {author} {\bibfnamefont {P.~A.}\ \bibnamefont
  {Lee}},\ }\Doi {10.1103/PhysRevB.57.R9381} {\bibfield  {journal} {\bibinfo
  {journal} {Phys. Rev. B},\ }\textbf {\bibinfo {volume} {57}},\ \bibinfo
  {pages} {R9381} (\bibinfo {year} {1998})}\BibitemShut {NoStop}%
\bibitem [{\citenamefont {Dmitriev}\ \emph {et~al.}(1997)\citenamefont
  {Dmitriev}, \citenamefont {Kachorovskii},\ and\ \citenamefont
  {Gornyi}}]{Dmit97}%
  \BibitemOpen
  \bibfield  {author} {\bibinfo {author} {\bibfnamefont {A.~P.}\ \bibnamefont
  {Dmitriev}}, \bibinfo {author} {\bibfnamefont {V.~Y.}\ \bibnamefont
  {Kachorovskii}}, \ and\ \bibinfo {author} {\bibfnamefont {I.~V.}\
  \bibnamefont {Gornyi}},\ }\Doi {10.1103/PhysRevB.56.9910} {\bibfield
  {journal} {\bibinfo  {journal} {Phys. Rev. B},\ }\textbf {\bibinfo {volume}
  {56}},\ \bibinfo {pages} {9910} (\bibinfo {year} {1997})}\BibitemShut
  {NoStop}%
\bibitem [{\citenamefont {Hikami}\ \emph {et~al.}(1980)\citenamefont
    {Hikami},
  \citenamefont {Larkin},\ and\ \citenamefont {Nagaoka}}]{Hik80}%
  \BibitemOpen
  \bibfield  {author} {\bibinfo {author} {\bibfnamefont {S.}~\bibnamefont
  {Hikami}}, \bibinfo {author} {\bibfnamefont {A.~I.}\ \bibnamefont {Larkin}},
  \ and\ \bibinfo {author} {\bibfnamefont {Y.}~\bibnamefont {Nagaoka}},\
  }\href@noop {} {\bibfield  {journal} {\bibinfo  {journal} {Prog. Theor.
  Phys.},\ }\textbf {\bibinfo {volume} {63}},\ \bibinfo {pages} {707 }
  (\bibinfo {year} {1980})}\BibitemShut {NoStop}%
\bibitem [{\citenamefont {Wittmann}\ and\ \citenamefont
  {Schmid}(1987)}]{Wit87}%
  \BibitemOpen
  \bibfield  {author} {\bibinfo {author} {\bibfnamefont {H.-P.}\ \bibnamefont
  {Wittmann}}\ and\ \bibinfo {author} {\bibfnamefont {A.}~\bibnamefont
  {Schmid}},\ }\href@noop {} {\bibfield  {journal} {\bibinfo  {journal} {J. Low
  Temp. Phys},\ }\textbf {\bibinfo {volume} {69}},\ \bibinfo {pages} {131 }
  (\bibinfo {year} {1987})}\BibitemShut {NoStop}%
\bibitem [{\citenamefont {Minkov}\ \emph {et~al.}(2004)\citenamefont
    {Minkov},
  \citenamefont {Germanenko},\ and\ \citenamefont {Gornyi}}]{Min04-2}%
  \BibitemOpen
  \bibfield  {author} {\bibinfo {author} {\bibfnamefont {G.~M.}\ \bibnamefont
  {Minkov}}, \bibinfo {author} {\bibfnamefont {A.~V.}\ \bibnamefont
  {Germanenko}}, \ and\ \bibinfo {author} {\bibfnamefont {I.~V.}\ \bibnamefont
  {Gornyi}},\ }\Doi {10.1103/PhysRevB.70.245423} {\bibfield  {journal}
  {\bibinfo  {journal} {Phys. Rev. B},\ }\textbf {\bibinfo {volume} {70}},\
  \bibinfo {pages} {245423} (\bibinfo {year} {2004})}\BibitemShut {NoStop}%
\bibitem [{\citenamefont {Minkov}\ \emph
  {et~al.}(2009){\natexlab{a}}\citenamefont {Minkov}, \citenamefont
  {Germanenko}, \citenamefont {Rut}, \citenamefont {Sherstobitov},\ and\
  \citenamefont {Zvonkov}}]{Min03}%
  \BibitemOpen
  \bibfield  {author} {\bibinfo {author} {\bibfnamefont {G.~M.}\ \bibnamefont
  {Minkov}}, \bibinfo {author} {\bibfnamefont {A.~V.}\ \bibnamefont
  {Germanenko}}, \bibinfo {author} {\bibfnamefont {O.~E.}\ \bibnamefont {Rut}},
  \bibinfo {author} {\bibfnamefont {A.~A.}\ \bibnamefont {Sherstobitov}}, \
  and\ \bibinfo {author} {\bibfnamefont {B.~N.}\ \bibnamefont {Zvonkov}},\
  }\Doi {10.1103/PhysRevB.79.235335} {\bibfield  {journal} {\bibinfo  {journal}
  {Phys. Rev. B},\ }\textbf {\bibinfo {volume} {79}},\ \bibinfo {pages}
  {235335} (\bibinfo {year} {2009}{\natexlab{a}})}\BibitemShut {NoStop}%
\bibitem [{\citenamefont {Finkel'stein}(1990)}]{FinRev}%
  \BibitemOpen
  \bibfield  {author} {\bibinfo {author} {\bibfnamefont {A.~M.}\ \bibnamefont
  {Finkel'stein}},\ }\enquote {\bibinfo {title} {Electron liquid in disordered
  conductors},}\ in\ \href@noop {} {\emph {\bibinfo {booktitle}
  {Electron-Electron Interaction in Disordered Systems}}},\ Vol.~\bibinfo
  {volume} {14},\ \bibinfo {editor} {edited by\ \bibinfo {editor}
  {\bibfnamefont {I.~M.}\ \bibnamefont {Khalatnikov}}}\ (\bibinfo  {publisher}
  {Harwood},\ \bibinfo {address} {London},\ \bibinfo {year} {1990})\BibitemShut
  {NoStop}%
\bibitem [{\citenamefont {Minkov}\ \emph
  {et~al.}(2009){\natexlab{b}}\citenamefont {Minkov}, \citenamefont
  {Germanenko}, \citenamefont {Rut}, \citenamefont {Sherstobitov},\ and\
  \citenamefont {Zvonkov}}]{Minkov09}%
  \BibitemOpen
  \bibfield  {author} {\bibinfo {author} {\bibfnamefont {G.~M.}\ \bibnamefont
  {Minkov}}, \bibinfo {author} {\bibfnamefont {A.~V.}\ \bibnamefont
  {Germanenko}}, \bibinfo {author} {\bibfnamefont {O.~E.}\ \bibnamefont {Rut}},
  \bibinfo {author} {\bibfnamefont {A.~A.}\ \bibnamefont {Sherstobitov}}, \
  and\ \bibinfo {author} {\bibfnamefont {B.~N.}\ \bibnamefont {Zvonkov}},\
  }\href@noop {} {\bibfield  {journal} {\bibinfo  {journal} {Phys. Rev. B},\
  }\textbf {\bibinfo {volume} {79}},\ \bibinfo {pages} {235335} (\bibinfo
  {year} {2009}{\natexlab{b}})}\BibitemShut {NoStop}%
\bibitem [{\citenamefont {Finkel'stein}(1984){\natexlab{b}}}]{Fin84}%
  \BibitemOpen
  \bibfield  {author} {\bibinfo {author} {\bibfnamefont {A.~M.}\ \bibnamefont
  {Finkel'stein}},\ }\href@noop {} {\bibfield  {journal} {\bibinfo  {journal}
  {Zh. Eksp. Teor. Fiz.},\ }\textbf {\bibinfo {volume} {86}},\ \bibinfo {pages}
  {367} (\bibinfo {year} {1984}{\natexlab{b}})},\ \translation{Sov. Phys. JETP
  \textbf{59}, 212 (1984)}\BibitemShut {NoStop}%
\bibitem [{\citenamefont {Raimondi}\ \emph {et~al.}(1990)\citenamefont
  {Raimondi}, \citenamefont {Castellani},\ and\ \citenamefont
  {Di~Castro}}]{Raim90}%
  \BibitemOpen
  \bibfield  {author} {\bibinfo {author} {\bibfnamefont {R.}~\bibnamefont
  {Raimondi}}, \bibinfo {author} {\bibfnamefont {C.}~\bibnamefont
  {Castellani}}, \ and\ \bibinfo {author} {\bibfnamefont {C.}~\bibnamefont
  {Di~Castro}},\ }\Doi {10.1103/PhysRevB.42.4724} {\bibfield  {journal}
  {\bibinfo  {journal} {Phys. Rev. B},\ }\textbf {\bibinfo {volume} {42}},\
  \bibinfo {pages} {4724} (\bibinfo {year} {1990})}\BibitemShut {NoStop}%
\bibitem [{\citenamefont {Minkov}\ \emph {et~al.}(2007)\citenamefont
    {Minkov},
  \citenamefont {Germanenko}, \citenamefont {Rut}, \citenamefont
  {Sherstobitov},\ and\ \citenamefont {Zvonkov}}]{Minkov07}%
  \BibitemOpen
  \bibfield  {author} {\bibinfo {author} {\bibfnamefont {G.~M.}\ \bibnamefont
  {Minkov}}, \bibinfo {author} {\bibfnamefont {A.~V.}\ \bibnamefont
  {Germanenko}}, \bibinfo {author} {\bibfnamefont {O.~E.}\ \bibnamefont {Rut}},
  \bibinfo {author} {\bibfnamefont {A.~A.}\ \bibnamefont {Sherstobitov}}, \
  and\ \bibinfo {author} {\bibfnamefont {B.~N.}\ \bibnamefont {Zvonkov}},\
  }\Doi {10.1103/PhysRevB.76.165314} {\bibfield  {journal} {\bibinfo  {journal}
  {Phys. Rev. B},\ }\textbf {\bibinfo {volume} {76}},\ \bibinfo {pages}
  {165314} (\bibinfo {year} {2007})}\BibitemShut {NoStop}%
\bibitem [{\citenamefont {Hikami}(1981)}]{Hik81}%
  \BibitemOpen
  \bibfield  {author} {\bibinfo {author} {\bibfnamefont {S.}~\bibnamefont
  {Hikami}},\ }\Doi {10.1103/PhysRevB.24.2671} {\bibfield  {journal} {\bibinfo
  {journal} {Phys. Rev. B},\ }\textbf {\bibinfo {volume} {24}},\ \bibinfo
  {pages} {2671} (\bibinfo {year} {1981})}\BibitemShut {NoStop}%
\bibitem [{\citenamefont {Wegner}(1979)}]{Wegner79}%
  \BibitemOpen
  \bibfield  {author} {\bibinfo {author} {\bibfnamefont {F.}~\bibnamefont
  {Wegner}},\ }\href@noop {} {\bibfield  {journal} {\bibinfo  {journal} {Z.
  Phys. B: Condens. Matter},\ }\textbf {\bibinfo {volume} {35}},\ \bibinfo
  {pages} {207} (\bibinfo {year} {1979})}\BibitemShut {NoStop}%
\bibitem [{\citenamefont {Wegner}(1980)}]{Wegner80}%
  \BibitemOpen
  \bibfield  {author} {\bibinfo {author} {\bibfnamefont {F.}~\bibnamefont
  {Wegner}},\ }\href@noop {} {\bibfield  {journal} {\bibinfo  {journal} {Phys.
  Rep.},\ }\textbf {\bibinfo {volume} {67}},\ \bibinfo {pages} {15} (\bibinfo
  {year} {1980})}\BibitemShut {NoStop}%
\bibitem [{\citenamefont {Wegner}(1989)}]{Wegner89}%
  \BibitemOpen
  \bibfield  {author} {\bibinfo {author} {\bibfnamefont {F.}~\bibnamefont
  {Wegner}},\ }\href@noop {} {\bibfield  {journal} {\bibinfo  {journal} {Nucl.
  Phys. B},\ }\textbf {\bibinfo {volume} {316}},\ \bibinfo {pages} {663}
  (\bibinfo {year} {1989})}\BibitemShut {NoStop}%
\bibitem [{\citenamefont {Baranov}\ \emph {et~al.}(2002)\citenamefont
  {Baranov}, \citenamefont {Burmistrov},\ and\ \citenamefont
  {Pruisken}}]{Baranov02}%
  \BibitemOpen
  \bibfield  {author} {\bibinfo {author} {\bibfnamefont {M.~A.}\ \bibnamefont
  {Baranov}}, \bibinfo {author} {\bibfnamefont {I.~S.}\ \bibnamefont
  {Burmistrov}}, \ and\ \bibinfo {author} {\bibfnamefont {A.~M.~M.}\
  \bibnamefont {Pruisken}},\ }\Doi {10.1103/PhysRevB.66.075317} {\bibfield
  {journal} {\bibinfo  {journal} {Phys. Rev. B},\ }\textbf {\bibinfo {volume}
  {66}},\ \bibinfo {pages} {075317} (\bibinfo {year} {2002})}\BibitemShut
  {NoStop}%
\bibitem [{\citenamefont {Punnoose}\ and\ \citenamefont
  {Finkel'stein}(2005)}]{Punnoose05}%
  \BibitemOpen
  \bibfield  {author} {\bibinfo {author} {\bibfnamefont {A.}~\bibnamefont
  {Punnoose}}\ and\ \bibinfo {author} {\bibfnamefont {A.~M.}\ \bibnamefont
  {Finkel'stein}},\ }\href@noop {} {\bibfield  {journal} {\bibinfo  {journal}
  {Science},\ }\textbf {\bibinfo {volume} {310}},\ \bibinfo {pages} {289}
  (\bibinfo {year} {2005})}\BibitemShut {NoStop}%
\bibitem [{\citenamefont {Shapiro}\ and\ \citenamefont
  {Abrahams}(1981)}]{Shapiro81}%
  \BibitemOpen
  \bibfield  {author} {\bibinfo {author} {\bibfnamefont {B.}~\bibnamefont
  {Shapiro}}\ and\ \bibinfo {author} {\bibfnamefont {E.}~\bibnamefont
  {Abrahams}},\ }\Doi {10.1103/PhysRevB.24.4025} {\bibfield  {journal}
  {\bibinfo  {journal} {Phys. Rev. B},\ }\textbf {\bibinfo {volume} {24}},\
  \bibinfo {pages} {4025} (\bibinfo {year} {1981})}\BibitemShut {NoStop}%
\bibitem [{\citenamefont {Gornyi}\ and\ \citenamefont
  {Novokshonov}()}]{PrvIgor}%
  \BibitemOpen
  \bibfield  {author} {\bibinfo {author} {\bibfnamefont {I.~V.}\ \bibnamefont
  {Gornyi}}\ and\ \bibinfo {author} {\bibfnamefont {S.~G.}\ \bibnamefont
  {Novokshonov}},\ }\href@noop {} {\bibinfo  {journal} {private
  communication}}\BibitemShut {NoStop}%
\end{thebibliography}
%merlin.mbs 2010-03-15 4.21a (PWD, AO, DPC)
\end{document}